\documentclass{article} 

\usepackage{subfiles}

\usepackage[final]{neurips_2019}

\usepackage[utf8]{inputenc} 
\usepackage[T1]{fontenc}    
\usepackage{url}
\usepackage{amsthm}
\usepackage{amsmath,amssymb,amsfonts}
\usepackage{hieroglf}
\usepackage{xcolor}
\usepackage{algorithm,algpseudocode}
\usepackage{tikz}
\usepackage{rbase} 
\usepackage{paralist}
\definecolor{hcitecolor}{RGB}{40,40,160}
\definecolor{hurlcolor}{RGB}{80,80,140}
\usepackage[colorlinks=true,citecolor=hcitecolor, urlcolor=hurlcolor, linkcolor=black]{hyperref}
\usepackage{booktabs}

\usepackage{tikz}
\usetikzlibrary{shadows}
\usetikzlibrary{matrix, calc, positioning, arrows,shapes,backgrounds, arrows.meta, quotes}
\usepackage{gensymb,stmaryrd,mathtools,textcomp,xspace,grffile,rbase}

\newtheorem{definition}{Definition}

\makeatletter
\def\thm@space@setup{%
  \thm@preskip=6pt plus 0pt minus 0pt
  \thm@postskip=0pt plus 0pt minus 0pt 
}
\makeatother

\newcommand{\neuron}{\mathfrak{n}}

\newcommand{\wt}{\widetilde}

\title{Cormorant:\m{\;} Covariant Molecular Neural Networks}

\author{Brandon Anderson\m{{}^{\ast\ddag}}, Truong-Son Hy\m{{}^\ast} and Risi Kondor\m{{}^{\ast\dag\sharp}}\\
\m{{}^{\ast}}Department of Computer Science, \m{{}^{\dag}}Department of Statistics\\
The University of Chicago\\
\m{{}^{\sharp}} Center for Computational Mathematics, Flatiron Institute\\
\m{{}^{\ddag}} Atomwise\\
\texttt{\{hytruongson,risi\}@uchicago.edu}\\
\texttt{brandona@jfi.uchicago.edu}
}

\begin{document}

\maketitle

\begin{abstract}
We propose \emph{Cormorant}, a rotationally covariant neural network architecture for learning the behavior 
and properties of complex many-body physical systems. 
We apply these networks to molecular systems with two goals: learning atomic potential energy surfaces for 
use in Molecular Dynamics simulations, and learning ground state properties of molecules calculated 
by Density Functional Theory. 
Some of the key features of our network are that (a) each neuron explicitly corresponds to a subset of atoms;  
(b) the activation of each neuron is covariant to rotations, ensuring that overall the network is fully rotationally invariant. 
Furthermore, the non-linearity in our network is based upon tensor products and the Clebsch-Gordan 
decomposition, allowing the network to operate entirely in Fourier space. 
\emph{Cormorant} significantly outperforms competing algorithms in learning molecular 
Potential Energy Surfaces from conformational geometries in the MD-17 dataset,  
and is competitive with other methods at learning geometric, energetic, electronic, and thermodynamic 
properties of molecules on the GDB-9 dataset.

\end{abstract}

\section{Introduction}

In principle, quantum mechanics provides a perfect description of the forces governing the 
behavior of atoms, molecules and crystalline materials such as metals.  
However, for systems larger than a few dozen atoms, solving  
the Schr\"odinger equation explicitly at every timestep 
is not a feasible proposition on present day computers. 
Even Density Functional Theory (DFT) \citep{HohenbergKohn}, 
a widely used approximation to the equations of quantum mechanics,  
has trouble scaling to more than a few hundred atoms. 

Consequently, the majority of practical work in molecular dynamics today  
falls back on fundamentally classical models, where the atoms are essentially treated as 
solid balls and the forces between them are given by pre-defined formulae called 
\emph{atomic force fields} or \emph{empirical potentials}, such as 
the CHARMM family of models \citep{CHARMM1983,CHARMM2009}.    
There has been a widespread realization that this approach 
has inherent limitations, so in 
recent years a burgeoning community has formed around trying to use machine learning 
to \emph{learn} more descriptive force fields directly from DFT computations 
\citep{Behler2007fe,Bartok2010wd,Rupp2012,Shapeev2015,Chmiela2016a,Zhang2017,Schutt2017,Hirn2017}.  
More broadly, there is considerable interest in using ML methods not just for learning  
force fields, but also for predicting many other physical/chemical properties of atomic systems 
across different branches of materials science, chemistry and pharmacology 
\citep{MontavonEtAl,Riley2017,ANI1,Tensormol}. 

At the same time, there have been significant advances in our understanding of the equivariance and 
covariance properties of neural networks, starting with 
\citep{Cohen2016,Cohen2017} in the context of traditional convolutional neural nets (CNNs). 
Similar ideas underly generalizations of CNNs to manifolds 
\citep{Masci2015,Monti2016,BronsteinEtAl} 
and graphs \citep{BrunaZaremba2014,HenaffLeCun2015}. 
In the context of CNNs on the sphere, \citet{SphericalCNN2018} realized the advantage of using 
``Fourier space'' activations, 
i.e., expressing the activations of neurons in a basis defined by the irreducible representations of the 
underlying symmetry group (see also \citep{EstevesSph}), and these ideas were later generalized 
to the entire \m{\textrm{SE}(3)} group \citep{Weiler}. 
\citet{EquivarianceICML18} gave a complete characterization of what operations are allowable in 
Fourier space neural networks to preserve covariance, and Cohen et al generalized the framework 
even further to arbitrary gauge fields \citep{CohenGauge}. 
There have also been some recent works 
where even the nonlinear part of the neural network's operation is performed in Fourier space: 
independently of each other \citep{Thomas2018} and \citep{Nbody2018arxiv} were to first to 
use the Clebsch--Gordan transform inside rotationally covariant neural networks for learning 
physical systems, while \citep{KLT2018} showed that in spherical CNNs the Clebsch--Gordan 
transform is sufficient to serve as the sole source of nonlinearity. 

The \emph{Cormorant} neural network architecture proposed in the present paper combines some of the 
insights gained from the various force field and potential learning efforts with the emerging theory of 
Fourier space covariant/equivariant neural networks. 
The important point that we stress in the following pages is that 
by setting up the network in such a way that each neuron 
corresponds to an actual set of physical atoms, and that each activation is covariant to symmetries 
(rotation and translation), we get a network in which the ``laws'' that individual 
neurons learn resemble known physical interactions. 
Our experiments show that this generality pays off in 
terms of performance on standard benchmark datasets. 

\ignore{
 
Assume that the potential attached to atom \m{i} is \m{\phi_i(\sseq{\h {\V r}}{k})},  
with \m{\h{\V r_j}=\V r_{p_j}\!\<-\V r_i}, where \m{\V r_i} is the position vector of atom \m{i} and 
\m{\V r_{p_j}} is the position vector of its \m{j}'th neighbor.  
The total force experienced by atom \m{i} is then simply the negative gradient  
\m{F_i=-\nabla_{\!\small \V r_i} \phi_i(\sseq{\h {\V r}}{k})}. 
Classically, in molecular dynamics \m{\phi_i} is usually given in terms of a closed form formula with a few tunable 
parameters. Popular examples of such so-called empirical potentials (empirical force fields) 
include the CHARMM models \citep{CHARMM1983,CHARMM2009} and others. 
 
Empirical potentials are fast to evaluate but are crude 
models of the quantum interactions between atoms, 
limiting the accuracy of molecular simulation. 
A little over ten years ago, machine learning entered this field, promising 
to bridge the gap between the quantum and classical worlds by \emph{learning} 
the aggregate force on each atom as a function of the positons of its neighbors from a relatively 
small number of DFT calculations \citep{Behler2007fe}. 
In the last few years there has been a veritable explosion in the amount of activity in 
machine learned atomic potentials (MLAP), 
and molecular dynamics simulations based on this approach 
are starting to yield results that outperform other methods  
\citep{Bartok2010wd,Behler2015gv,Shapeev2015,Chmiela2016a,Zhang2017,Schutt2017}.  

Much of the arsenal of present day machine learning algorithms has been applied to the MLAP problem,  
from genetic algorithms, through kernel methods, to neural networks. 
However, rather than the statistical 
details of the specific learning algorithm, often what is critically important for 
problems of this type 
is the representation of the atomic environment, i.e., the choice of learning features  
that the algorithm is based on. 
This situation is by no means unique in the world of applied machine learning:  
in computer vision and speech recognition, in particular, there is a rich literature 
of such representational issues. 
What makes the situation in Physics applications somewhat special is the presence of 
constraints and invariances that the representation must satisfy not just in an approximate, 
but in the \emph{exact} sense.  
As an example, one might consider rotation invariance. 
If rotation invariance is not fully respected by an image recognition system, some objects 
might be less likely to be accurately detected in certain orientations than in others. 
In a molecular dynamics setting, however, using a potential that is not fully rotationally invariant 
would not just degrade accuracy, but would likely lead to entirely unphysical molecular trajectories.  

\subsection{Fixed vs.\:learned representations.} 
Similarly to other branches of machine learning, in recent years the MLAP community has been shifting from 
fixed input features towards representations learned from the data itself, 
in particular, using ``deep'' neural networks to represent atomic enviroments. 
Several authors have found that certain concepts from the mainstream neural networks literature,  
such as convolution and equivariance, can be successfuly repurposed to this domain. 
In fact, the analogy with computer vision is more than just skin deep. 
In both domains two competing objectives are critical to success: 
\begin{compactenum}[~~1.]
\item 
The ability to capture structure in the input data at multiple different length scales, 
, i.e., to construct a \emph{multiscale} representation of the input image or the atomic environment. 
\item 
The above mentioned invariance property with respect to spatial transformations, including  
translations, rotations, and possibly scaling. 
\end{compactenum}
There is a rich body of work on addressing these objectives in the neural networks 
literature. One particularly attractive approach is 
the \emph{scattering networks} framework of Mallat and coworkers, 
which, at least in the limit of an infinite number of neural network layers, 
provides a representation of functions that is both 
globally invariant with respect to symmetries and Lipschitz with respect to warpings 
\citep{Mallat2012,Hirn2017}. 

Inspired by recent work on neural networks for representing graphs and other structured objects 
by covariant compositional neural architectures \citep{CompNetsArxiv18}, 
in this paper we take the idea of learnable multiscale representations one step further,  
and propose \m{N}--body networks, 
a neural network architecture  where \emph{the individual ``neurons'' 
correspond to physical subsystems endowed with their own internal state}.
The structure and behavior of the resulting model 
follows the tradition of coarse graining and representation theoretic ideas in Physics, and 
provides a learnable and multiscale representation of the atomic environment that is fully 
covariant to the action of the appropriate symmetries. 
However, the scope of the underlying ideas is significantly broader, and we believe that \m{N}--body networks 
will also find application in modeling other types of many-body Physical systems, as well. 

An even more general contribution of the present work is that it shows how the machinery of group 
representation theory, specifically the concept of Clebsch--Gordan decompositions, can be used to design 
neural networks that are covariant to the action of a compact group yet are computationally efficient. 
This aspect is related to the recent explosion of interest in generalizing the notion of convolutions to graphs 
\citep{Niepert2016,Defferrard2016,Duvenaud2015,Li2016,Riley2017,CompNetsArxiv18}, 
manifolds \citep{Monti2016, Masci2015}, 
and other domains \citep{Bruna2013,SphericalCNN2018},  
as well as the question of generalizing the concept of equivariance (covariance) in general   
\citep{Cohen2016,Cohen2017,EquivarianceArxiv18}. 
Several of the above works employed generalized Fourier representations of one type or another, 
but to ensure equivariance the nonlinearity was always applied in the ``time domain''. Projecting 
back and forth between the time domain and the frequency domain is a major bottleneck, 
which we can eliminate because the Clebsch--Gordan 
transform allows us to compute one type of nonlinearity, tensor products, entirely in the Fourier domain.  
}


\section{The nature of physical interactions in molecules}\label{sec: physical}

Ultimately interactions in molecular systems arise from the
quantum structure of electron clouds around constituent atoms.
However, from a chemical point of view, effective atom-atom interactions
break down into a few simple classes based upon symmetry. 
Here we review a few of these classes in the context of the multipole expansion, whose structure will inform the design of our neural network. 


\paragraph{Scalar interactions.}

The simplest type of physical interaction is that between two particles that are pointlike and 
have no internal directional degrees of freedom, such as spin or dipole moments. A classical example is 
the electrostatic attraction/repulsion between two charges described by the Coulomb energy 
\begin{equation}\label{eq: Coulomb}
V_C
=-\ovr{4\pi\epsilon_0}\,\fr{q_A q_B}{\absN{\V r_{\!AB}}}\:. 
\end{equation}
Here \m{q_A} and \m{q_B} are the charges of the two particles, \m{\V r_{\!A}} and \m{\V r_{\nts B}} are 
their position vectors, \m{\V r_{\!AB}=\V r_{\!A}\<-\V r_{\nts B}}, and \m{\epsilon_0} is a universal constant. 
Note that this equation already reflects symmetries: the fact that \rf{eq: Coulomb} only depends on the 
\emph{length} of \m{\V r_{\!AB}} and not its direction or the position vectors individually 
guarantees that the potential is invariant under both translations and rotations.  

\paragraph{Dipole/dipole interactions.}

One step up from the scalar case is the interaction between two dipoles. 
In general, the electrostatic dipole moment of a set of \m{N} charged particles relative 
to their center of mass \m{\V r} is just the first moment of their position vectors weighted by 
their charges:
\[\V \mu=\sum_{i=1}^{N}q_i (\V r_i-\V r).\]
The dipole/dipole contribution to the electrostatic potential energy between two sets of particles 
\m{A} and \m{B} separated by a vector \m{\V r_{\!AB}} is then given by 
\begin{equation}\label{eq: dipole-dipole}
V_{d/d}=\ovr{4\pi\epsilon_0}\sqbbigg{\fr{\V\mu_A\cdot\V\mu_B}{\abs{\V r_{\!AB}}^3}-3\,
\fr{(\V \mu_A\cdot\V r_{\!AB})(\V \mu_B\cdot\V r_{\!AB})}{|\V r_{\!AB}|^5}}.
\end{equation}
One reason why dipole/dipole interactions are indispensible for capturing the energetics of molecules 
is that most chemical bonds are polarized. However, dipole/dipole interactions also occur 
in other contexts, such as the interaction between the magnetic spins of electrons. 

\paragraph{Quadropole/quadropole interactions.}

One more step up the multipole hierarchy is the interaction between quadropole moments. 
In the electrostatic case, the quadropole moment is the second moment of the charge density 
(corrected to remove the trace), described by the \emph{matrix} 
\[\V \Theta=\sum_{i=1}^N q_i(3\tts \V r_i\V r_i^\top-\abs{\V r_i}^2\nts I).\] 
Quadropole/quadropole interactions appear for example when describing the interaction between benzene rings, 
but the general formula for the corresponding potential is quite complicated. 
As a simplification, 
let us only consider the special case when in some coordinate system aligned with the structure of 
\m{A}, and at polar angle \m{(\theta_A,\phi_A)} relative to the vector \m{\V r_{\!AB}} connecting \m{A} and \m{B}, 
\m{\V \Theta_A} can be transformed into a form that is diagonal, with \m{[\Theta_{\nts A}]_{zz}\<=\vartheta_A} 
and \sm{[\Theta_{\nts A}]_{xx}\<=[\Theta_{\nts A}]_{yy}\<=-\vartheta_A/2} \citep{stone1997theory}. 
We make a similar assumption about the quadropole moment of \m{B}.  
In this case the interaction energy becomes 
\begin{multline}\label{eq: quad-quad}
V_{q/q}=\fr{3}{4}\fr{\vartheta_A\vartheta_B}{4\pi\epsilon_0 \abs{\V r_{\!AB}}^5}
\big[ 1-5\cos^\theta_A-5\cos^2\theta_B-15\cos^2\theta_A\cos^2\theta_B+\\2(4\cos\theta_A\theta_B-
\sin\theta_A\sin\theta_B\cos(\phi_A\<-\phi_B))^2\big].
\end{multline}
Higher order interactions involve moment tensors of order 3,4,5, and so on. 
One can appreciate that the corresponding formulae, especially when considering not just electrostatics 
but other types of interactions as well (dispersion, exchange interaction, etc), 
quickly become very involved. 

\section{Spherical tensors and representation theory}

Fortunately, there is an alternative formalism for expressing molecular interactions, that of 
spherical tensors, which makes the general form of physically allowable 
interactions more transparent. 
This formalism also forms the basis of the  
our Cormorant networks described in the next section. 

The key to spherical tensors is understanding how physical quantities transform under rotations.  
Specifically, in our case, under a rotation \m{\V R}: 
\[q\longmapsto q\hspace{40pt}
\V\mu\longmapsto \V R\ts\V\mu\hspace{40pt}
\V \Theta\longmapsto \V R\ts \V \Theta \V R^\top\hspace{40pt}
\V r_{\!AB} \longmapsto \V R\,\V r_{\!AB}.
\]
Flattening \sm{\V\Theta} into a vector \sm{\wbar{\V\Theta}\tin\RR^9}, its transformation rule 
can equivalently be written as \sm{\wbar{\V\Theta}\mapsto (\V R\<\otimes \V R)\, \wbar{\V\Theta}}, 
showing its similarity to the other three cases.  
In general, a \m{k}'th order Cartesian moment tensor \sm{T^{(k)}\tin \RR^{3\times 3\times \ldots \times 3}} (or its flattened \sm{\wbar T{}^{(k)} \in \RR^{3k}} equivalent)
transforms as 
\sm{\wbar T{}^{(k)}\mapsto (\V R\<\otimes\V R\<\otimes\ldots\otimes\V R)\,\wbar T{}^{(k)}}.  

Recall that given a group \m{G}, a \emph{representation} \m{\rho} of \m{G} is  
a matrix valued function \m{\rho\colon G\to\CC^{d\times d}} obeying 
\m{\rho(xy)=\rho(x)\rho(y)} for any two group elements \m{x,y\tin G}. 
It is easy to see that \m{\V R}, and consequently \m{\V R\otimes\ldots\otimes\V R} are representations 
of the three dimensional rotation group \m{\SO(3)}.  
We also know that because \m{\SO(3)} is a compact group, it has a countable sequence of unitary  
so-called irreducible representations (irreps), and,  
up to a similarity transformation, any representation can be reduced to a direct sum of irreps. 
In the specific case of \m{\SO(3)}, the irreps are called \emph{Wigner D-matrices} and 
for any positive integer \m{\ell=0,1,2,\ldots} there is a single corresponding irrep \sm{D^\ell(\V R)}, 
which is a \m{(2\ell\<+1)} dimensional representation  
(i.e., as a function, \sm{D^\ell\colon \SO(3)\to \CC^{(2\ell+1)\times(2\ell+1)}}).  
The \m{\ell\<=0} irrep is the trivial irrep \m{D^0(\V R)\<=(1)}. 

The above imply that there is a fixed unitary transformation matrix \m{C^{(k)}} which reduces 
the \m{k}'th order rotation operator 
into a direct sum of irreducible representations:     
\[\underbrace{\V R\<\otimes\V R\<\otimes\ldots\otimes\V R}_k=C^{(k)}
\sqbBig{\bigoplus_\ell \bigoplus_{i=1}^{\tau_\ell} D^\ell(\V R)} {C^{(k)}}^\dag. 
\]
Note that the transformation \sm{\V R\<\otimes\V R\<\otimes\ldots\otimes\V R} contains redundant copies of \m{D^\ell(\V R)}, which we denote as the multiplicites \m{\tau_\ell}. For our present purposes knowing the actual values of the \m{\tau_\ell} 
is not that important, except that \m{\tau_k\<=1} and that 
for any \m{\ell>k},~ \m{\tau_\ell\<=0}. 
What is important is that \sm{\wbar T{}^{(k)}}, the vectorized form of the Cartesian moment tensor has a 
corresponding decomposition \vspace{-6pt}
\begin{equation}\label{eq: Cartesian decomp}
\wbar T{}^{(k)}=C^{(k)} \sqbBig{\bigoplus_\ell \bigoplus_{i=1}^{\tau_\ell} Q_{\ell,i}}. 
\end{equation}
This is nice, because using the unitarity of \sm{Q_{\ell_i}}, it shows that 
under rotations the individual \sm{Q_{\ell,i}}  components transform \emph{independently}  
as \sm{Q_{\ell,i}\mapsto D^\ell(\V R)\ts Q_{\ell,i}}. 

What we have just described is a form of generalized Fourier analysis applied to 
the transformation of Cartesian tensors under rotations. 
For the electrostatic multipole problem it is particularly relevant, because it turns out 
that in that case, due to symmetries of \sm{\wbar{T}{}^{(k)}}, 
the only nonzero \sm{Q_{\ell,i}} component of \rf{eq: Cartesian decomp} is the single one with \m{\ell\<=k}. 
Furthermore, for a set of \m{N} charged particles (indexing its components \m{-\ell,\ldots,\ell}) 
\m{Q_\ell} has the simple form  
\begin{equation}
[Q_\ell]_m=\br{\fr{4\pi}{2\ell\<+1}}^{1/2}\sum_{i=1}^{N} q_i\, (r_i)^\ell\:Y_\ell^m(\theta_i,\phi_i)
\hspace{60pt}m=-\ell,\ldots,\ell,\label{eq:multipole-aggregation}
\end{equation}
where \m{(r_i,\theta_i,\phi_i)} are the coordinates of the \m{i}'th particle in 
spherical polars, and the \m{Y_\ell^m(\theta,\phi)} are the well known spherical harmonic functions. 
\m{Q_\ell} is called the \m{\ell}'th \emph{spherical moment} of the charge distribution. 
Note that while \sm{\wbar{T}{}^{(\ell)}} and \sm{Q_\ell} convey exactly the same information, 
\sm{\wbar{T}{}^{(\ell)}} is a tensor with \sm{3^{\ell}} components, while \m{Q_\ell} is just a 
\m{(2\ell\<+1)} dimensional vector. 

Somewhat confusingly, in physics and chemistry any quantity \m{U} that transforms under rotations as 
\sm{U\<\mapsto D^\ell(\V R)\ts U} is often called an (\m{\ell}'th order) \emph{spherical tensor}, 
despite the fact that in terms of its presentation \m{Q_\ell} is just a vector of \m{2\ell\<+1} numbers. 
Also note that since \m{D^0(\V R)\<=(1)}, a zeroth order spherical tensor is just a scalar. 
A first order spherical tensor, on the other hand, can be used to represent a spatial vector 
\m{\V r\<=(r,\theta,\phi)} by setting \m{[U_1]_m=r\, Y_1^m(\theta,\phi)}. 

\subsection{The general form of interactions}
\label{sec: gen interaction}

The benefit of the spherical tensor formalism is that it makes it very clear how each part of 
a given physical equation transforms under rotations. For example, if \m{Q_\ell} and \sm{\wt Q_\ell} are two 
\m{\ell}'th order spherical tensors, then \sm{Q_\ell^\dag  \wt Q_\ell} is a scalar, 
since under a rotation \m{\V R}, by the unitarity of the Wigner \m{D}-matrices, 
\[Q_\ell^\dag  \wt Q_\ell\longmapsto (D^\ell\nts(\V R)\, Q_\ell)^\dag\, (D^\ell\nts(\V R)\, \wt Q_\ell)=
 Q_\ell^\dag\: (D^\ell\nts(\V R))^\dag \, D^\ell\nts(\V R)\: \wt Q_\ell=Q_\ell^\dag  \wt Q_\ell. 
\]
Even the dipole/dipole interaction \rf{eq: dipole-dipole} 
requires a more sophisticated way of coupling spherical 
tensors than this, since it involves non-trivial interactions between not just two, but three different quantites: 
the two dipole moments \sm{\V \mu_{\!A}} and \sm{\V \mu_B} and the the relative position vector 
\sm{\V r_{\!AB}}. 
Representing interactions of this type requires taking \emph{tensor products} of the constituent variables. 
For example, in the dipole/dipole case we need terms of the form \sm{Q_{\ell_1}^A\otimes Q_{\ell_2}^B}. 
Naturally, these will transform according to the tensor product of the corresponding irreps: 
\[Q_{\ell_1}^A\<\otimes Q_{\ell_2}^B\mapsto (D^{\ell_1}\!(\V R)\<\otimes D^{\ell_2}\!(\V R))\, 
(Q_{\ell_1}^A\<\otimes Q_{\ell_2}^B).\]
In general, \sm{D^{\ell_1\!}(\V R)\<\otimes D^{\ell_2\!}(\V R)} is \emph{not} an irreducible representation.  
However it does have a well studied decomposition into irreducibles, called the \emph{Clebsch--Gordan} 
decomposition: 
\[D^{\ell_1\!}(\V R)\<\otimes D^{\ell_2\!}(\V R)=C_{\ell_1,\ell_2}^\dag
\sqbbigg{\:\bigoplus_{\ell=\abs{\ell_1-\ell_2}}^{\ell_1+\ell_2} D^\ell\nts(\V R)\,} C_{\ell_1,\ell_2}.\]
Letting \m{C_{\ell_1,\ell_2,\ell}\tin \CC^{(2\ell+1)\times(2\ell_1+1)(2\ell_2+2)}} 
be the block of \m{2\ell\<+1} rows in \m{C_{\ell_1,\ell_2}} corresponding to the \m{\ell} component of 
the direct sum, we see that 
\m{C_{\ell_1,\ell_2,\ell}(Q_{\ell_1}^A\<\otimes Q_{\ell_2}^B)} is an \m{\ell}'th order spherical tensor. 
In particular, given some other spherical tensor quantity \sm{U_\ell}, 
\[U_\ell^\dag \cdot  C_{\ell_1,\ell_2,\ell}\cdot (Q_{\ell_1}^A\<\otimes Q_{\ell_2}^B)\]
is a scalar, and hence it is a candidate for being a term in the potential energy. 
Note the similarity of this expression to the \emph{bispectrum} \citep{KakaralaPhD,Bendory}, 
which is an already established tool in the force field learning literature \citep{BartokPRB2013}. 

Almost any rotation invariant interaction potential can be expressed in terms of iterated Clebsch--Gordan 
products between spherical tensors. 
In particular, the full electrostatic energy between two sets of charges \m{A} and \m{B} 
separated by a vector \m{\V r=(r,\theta,\phi)} expressed in multipole form \citep{jackson_classical_1999} is  
\begin{equation}
V_{AB}=\ovr{4\pi\epsilon_0} \sum_{\ell=0}^\infty \sum_{\ell'=0}^\infty 
\sqrt{{2\ell+2\ell'}\choose{2\ell}}
\sqrt{\fr{4\pi}{2\ell\<+2\ell'+1}}\: 
r^{-(\ell+\ell'+1)}\:
 Y_{\ell+\ell'}(\theta,\phi)\, C_{\ell_1,\ell_2,{\ell+\ell'}}\:(Q^A_{\ell} \otimes Q^B_{\ell'}). \label{eq:multipole-energy}
\end{equation}
Note the generality of this formula: the \m{\ell\<=\ell'\<=1} case covers the dipole/dipole 
interaction \rf{eq: dipole-dipole}, the \m{\ell\<=\ell'\<=2} case covers the quadropole/quadropole 
interaction \rf{eq: quad-quad}, while the other terms cover every other possible type of 
multipole/multipole interaction. 
Magnetic and other types of interactions, including interactions that involve  
3-way or higher order terms, can also be 
recovered from appropriate combinations of tensor products and Clebsch--Gordan decompositions. 

We emphasize that our discussion of electrostatics is only intended to illustrate the algebraic structure of interatomic interactions of any type, and is not restricted to electrostatics. In what follows, we will not explicitly specify what interactions the network will learn. Nevertheless, there are physical constraints on the interactions arising from symmetries, which we explicitly impose in our design of Cormorant.

\ignore{
\[D^\ell\colon \SO(3)\to \CC^{(2\ell+1)\times(2\ell+1)}.\]
Also recall that the \emph{spherical harmonics} 
\m{\setofN{Y_\ell^m(\theta,\phi)}{\ell\<=0,1,2,\ldots,~m\<=-\ell,\ldots,\ell}} 
are a basis for complex functions on the unit sphere \m{S^2}. 
The fact that the number of 
spherical harmonics for a given value of \m{\ell} is the same as the dimensionality of \m{D^\ell} is 
not a coincidence: the Wigner D-matrices and the spherical harmonics are closely related, in particular 
\m{f} is a function on \m{S^2} and 
\[f(\theta,\phi)=\sum_{l=0}^{\infty} \sum_{m=-\ell}^{\ell}\h f^\ell_m Y^\ell_m(\theta,\phi),\]
is its spherical harmonic expansion, then if 
we apply a rotation \m{\V R} to \m{f}, the expansion coefficients for any given \m{\ell}, 
collected in a vector \m{\h f^\ell}, change to 
\[\h {\V f}'^\ell=D^\ell(\V R)\,\h {\V f}^\ell.\]
}
\newcommand{\otimescg}{\otimes_{\rm cg}}

\section{CORMORANT:\m{\:} COvaRiant MOleculaR Artificial Neural neTworks}\label{sec: cormorant}

The goal of using ML in molecular problems is not to encode known physical laws, 
but to provide a platform for learning interactions from data that cannot easily be captured in a 
simple formula. 
Nonetheless, the mathematical structure of known physical laws, like those discussed in the previous sections, 
give strong hints about how to represent physical interactions in algorithms.  
In particular, when using machine learning to learn molecular potentials or similar 
rotation and translation invariant physical quantities, it is essential to make sure that the algorithm 
respects these invariances. 

Our Cormorant neural network has invariance to rotations baked into its architecture in a way that 
is similar to the physical equations of the previous section:  
the internal activations are all spherical tensors, which are then 
combined at the top of the network in such a way as to guarantee that the final output is a scalar 
(i.e., is invariant). 
However, to 
allow the network to learn interactions that are more complicated than classical interatomic  
forces, we allow each neuron to output not just a single spherical tensor, but a combination 
of spherical tensors of different orders. 
We will call an object consisting of \sm{\tau_0} scalar components, \m{\tau_1} components transforming as first order 
spherical tensors, \m{\tau_2} components transforming as second order spherical tensors, and so on,  
an \m{\SO(3)}--\emph{covariant vector of type} \sm{(\tau_0,\tau_1,\tau_2,\ldots)}. 
The output of each neuron in Cormorant is an \m{\SO(3)}--vector of a fixed type. 


\begin{definition}
We say that \m{F} is an \m{\SO(3)}-covariant vector of type 
\sm{\V \tau=(\tau_0,\tau_1,\tau_2,\ldots,\tau_L)} if it can be written as a collection of 
complex matrices \sm{F_0,F_1,\ldots,F_L}, called its \emph{isotypic parts},   
where each \m{F_\ell} is a matrix of size \m{(2\ell\<+1) \<\times \tau_\ell} and transforms under rotations 
as \m{F_\ell\mapsto D^\ell(\V R)\, F_\ell}. 
\end{definition}
 
The second important feature of our architecture is that each neuron corresponds to 
either a single atom or a set of atoms forming a physically meaningful subset of the system at hand, 
for example all atoms in a ball of a given radius. 
This condition helps encourage the network to learn physically meaningful and interpretable 
interactions. 
The high level definition of \emph{Cormorant} nets is as follows.

\begin{definition} \label{def:cormorant}
Let \m{\Scal} be a molecule or other physical system consisting of \m{N} atoms.
A ``Cormorant'' covariant molecular neural network for \m{\Scal} is a feed forward 
neural network consisting of \m{m} neurons \m{\sseq{\neuron}{m}}, such that  
\vspace{-4pt}
\begin{enumerate}[~~C1.]
\item Every neuron \m{\neuron_i} corresponds to some subset \m{\Scal_i} of the atoms. 
In particular, each input neuron corresponds to a single atom. Each output neuron 
corresponds to the entire system \m{\Scal}. 
\item The activation of each \m{\neuron_i} is an \m{\SO(3)}-vector of a fixed type 
\m{\V \tau_{\!i}}.
\item The type of each output neuron is \sm{\V \tau_{\!\textrm{out}}\<=(1)}, i.e., a scalar.~\footnote{Cormorant can learn data of arbitrary $\textrm{SO(3)}$-vector outputs. We restrict to scalars here to simplify the exposition.}
\end{enumerate}
\end{definition} 

Condition (C3) guarantees that whatever function a \emph{Cormorant} network 
learns will be invariant to global rotations. Translation invariance is 
easier to enforce simply by making sure that the interactions represented by individual neurons 
only involve relative distances. 

\subsection{Covariant neurons}
\label{sec:cormorant_neurons}

The neurons in our network must be such that if each of their inputs is 
an \m{\SO(3)}--covariant vector then so is their output. 
Classically, neurons perform a simple linear operation such as \m{\x\mapsto W\x+\V b}, 
followed by a nonlinearity like a ReLU. 
In convolutional neural nets the weights are tied together in a specific way which 
guarantees that the activation of each layer is covariant to the action of global translations. 
\citet{EquivarianceICML18} discuss the generalization of convolution to the action of compact groups 
(such as, in our case, rotations) and prove that the only possible \emph{linear} operation 
that is covariant with the group action, is what, in terms of \m{\SO(3)}--vectors, corresponds to 
multiplying each \m{F_\ell} matrix from the right by some matrix \m{W} of learnable weights. 

For the nonlinearity, one option would be to express each spherical tensor as a function on 
\m{\SO(3)} using an inverse \m{\SO(3)} Fourier transform, apply a pointwise nonlinearity, and then 
transform the resulting function back into spherical tensors. This is the approach taken in 
e.g., \citep{SphericalCNN2018}. However, in our case this would be forbiddingly 
costly, as well as introducing quadrature errors by virtual of having to interpolate on the group, 
ultimately degrading the network's covariance. 
Instead, taking yet another hint from the structure of physical interactions, we use 
the Clebsch--Gordan transform introduced in \ref{sec: gen interaction} as a nonlinearity. 
The general rule for taking the CG product of two \m{\SO(3)}--parts 
\sm{F_{\ell_1}\tin\CC^{(2\ell_1+1)\times n_1}} 
and \sm{G_{\ell_2}\tin \CC^{(2\ell_2+1)\times n_2}} gives a collection of 
parts \sm{[F_{\ell_1}\otimescg G_{\ell_2}]_{\abs{\ell_1-\ell_1}},\ldots [F_{\ell_1}\otimescg G_{\ell_2}]_{\ell_1+\ell_1}} 
with columns 
\begin{equation}\label{CG parts}
\sqbbig{[F_{\ell_1}\otimescg G_{\ell_2}]_\ell}_{\ast,(i_1,i_2)}=
C_{\ell_1,\ell_2,\ell} \br{[F_{\ell_1}]_{\ast,i_1} \otimes [G_{\ell_2}]_{\ast,i_2}},
\end{equation}
i.e., every column of \sm{F_{\ell_1}} is separately CG-multiplied with every column of \sm{G_{\ell_2}}.  
The \m{\ell}'th part of the CG-product of two \m{\SO(3)}--vectors consists of the concatenation of all 
\m{\SO(3)}--part matrices with index \m{\ell} coming from multiplying each part of \m{F} 
with each part of \m{G}:
\[ [F\otimescg G]_\ell=\bigoplus_{\ell_1} \bigoplus_{\ell_2}  [F_{\ell_1} \otimescg G_{\ell_2}]_\ell.\]
Here and in the following \m{\oplus} denotes the appropriate concatenation of vectors and matrices. 
In Cormorant, however, as a slight departure from \rf{CG parts}, 
to reduce the quadratic blow-up in the number of columns, we always have \m{n_1\<=n_2} 
and use the restricted ``channel-wise'' CG-product, 
\[
\sqbbig{[F_{\ell_1}\otimescg G_{\ell_2}]_\ell}_{\ast,i}=
C_{\ell_1,\ell_2,\ell} \br{[F_{\ell_1}]_{\ast,i} \otimes [G_{\ell_2}]_{\ast,i}},
\]
where each column of \m{F_{\ell_1}} is only mixed with the corresponding column of \m{G_{\ell_2}}. 
We note that similar Clebsch--Gordan nonlinearities were used in \citep{KLT2018}, and that 
the Clebsch--Gordan product is  also an essential part of Tensor Field Networks \citep{Thomas2018}.


\subsection{One-body and two-body interactions}
\label{sec:n_atom_interactions}

As stated in Definition \ref{def:cormorant}, the covariant neurons in a Cormorant net correspond to different 
subsets of the atoms making up the physical system to be modeled. For simplicty in our present architecture 
there are only two types of neurons: those that correspond to individual atoms and those that correspond 
to pairs. For a molecule consisting of \m{N} atoms, each layer \m{s=0,1,\ldots, S} of the covariant 
part of the network has \m{N} neurons corresponding to the atoms and \m{N^2} neurons corresponding 
to the \m{(i,j)} atom pairs. By loose analogy with graph neural networks, we call the corresponding 
\m{F_i^s} and \m{g^s_{i,j}} activations vertex and edge activations, respectively. 

In accordance with the foregoing, each \m{F_i^s} 
activation is an \m{\SO(3)}--vector consisting of \m{L\<+1} 
distinct parts \sm{(F_i^{s,0},F_i^{s,1},\ldots, F_i^{s,L})},  
i.e., each \m{F_i^{s,\ell}} is a 
\m{(2\ell+1)\times \tau^s_\ell} dimensional complex matrix that transforms under rotations as 
\m{F_i^{s,\ell}\mapsto D^\ell(R)\, F_i^{s,\ell}}. 
The different columns of these matrices are regarded as the different \emph{channels} of the network, 
because they fulfill a similar role to channels in conventional convolutional nets. 
The \m{g^{s}_{i,j}} edge activations also break down into parts 
\sm{(g_{i,j}^{s,0},g_{i,j}^{s,1},\ldots, g_{i,j}^{s,L})}, but these are invariant under rotations. 
Again for simplicity, in the version of Cormorant that we used in our experiments \m{L} is the same 
in every layer (specifically \m{L\<=3}), and the number of channels is also 
independent of both \m{s} and \m{\ell}, specifically, \m{\tau^s_\ell\<\equiv n_c\<=16}.

The actual form of the vertex activations captures ``one-body interactions'' 
propagating information from the previous layer related to the \emph{same} atom and  
(indirectly, via the edge activations) ``two-body interactions'' 
capturing interactions between \emph{pairs} of atoms: 
\begin{equation}\label{eq: vertex activations}
F^{s-1}_{i}= 
\Big[
\underbrace{F^s_i \oplus \big(F^{s-1}_{i}\otimescg F^{s-1}_{i}\big)}_{\text{one-body part}} \oplus 
\underbrace{\Big( \sum_{j} G_{i,j}^{s} \otimescg F^{s-1}_{j} \Big)}_{\text{two-body part}}\Big]\cdot 
W^{\text{vertex}}_{s,\ell}.\\
\end{equation}
Here \m{G_{i,j}^{s}} are \m{\SO(3)}--vectors arising from the edge network. 
Specifically, \m{G_{i,j}^{s,\ell}=g_{i,j}^{s,\ell}\, Y^\ell(\V {\h r}_{i,j})}, 
where \m{Y^\ell(\V {\h r}_{i,j})} are the spherical harmonic 
vectors capturing  the relative position of atoms \m{i} and \m{j}. 
The edge activations, in turn, are defined  
\begin{equation}
g^{s,\ell}_{i,j}={\,\mu^s_{}(r_{i,j})\: \sqbBig{\brbig{\,g^{s-1,\ell}_{i,j}\oplus 
\brbig{F^{s-1}_{i}\cdot F^{s-1}_{j}}\oplus 
\eta^{s,\ell}_{}(r_{i,j})\,} \, W^{\text{edge}}_{s,\ell}\,}} 
\end{equation}
where we made the \m{\ell=0,1,\ldots,L} irrep index explicit. 
As before, in these formulae, \m{\oplus} denotes concatenation over the channel index \m{c},  
\m{\eta^{s,\ell}_{c}(r_{i,j})} 
are learnable radial functions, and \sm{\mu^s_{c}(r_{i,j})} are learnable cutoff functions limiting the 
influence of atoms that are farther away from atom \m{i}. 
The learnable parameters of the network are the \m{\cbrN{W^{\text{vertex}}_{s,\ell}}} and 
\m{\cbrN{W^{\text{edge}}_{s,\ell}}} weight matrices. 

Note that the \sm{F^{s-1}_{i}\cdot F^{s-1}_{j}} dot product term is the only term 
in these formulae responsible for the interaction between different atoms, and that this term always appears in conjunction with the 
\m{\eta^{s,\ell}_{c}(r_{i,j})} radial basis functions and \sm{\mu^s_{c}(r_{i,j})} cutoff functions 
(as well as the \m{\SO(3)}--covariant spherical harmonic vector) making sure that interaction scales 
with the distance between the atoms. 
More details of these activation rules are given in the Supplement. 

\ignore{
In order to motivate our Cormorant architecture, let's return to the multipole problem above. 
We start with Eq.~\rf{eq:multipole-aggregation}, which demonstrates how to construct a single moment from a set of charges. 
We denote this operation a ``one-body'' term, as it creates a new representation from a linear operation over 
input representations (in this case, the charges $q_i$.) On the other hand, the electrostatic energy in 
Eq.~\rf{eq:multipole-energy} is a ``two-body'' term, as it arises from the interaction between two different input representations, $Q^{A}_{\ell}$ and $Q^{B}_{\ell^\prime}$. 

We generalize this analogy to propose a framework for designing our Cormorant: 
we define an $n$-atom term as the Clebsch-Gordan product of $n$ $\SO(3)$-vector activations, 
which we henceforth denote as $F_i$ (instead of $Q_i$ as above) following the convention in neural networks.
For the purposes of this manuscript we start by considering activations that have the form
\begin{equation}
F^{s+1}_{i} = \Big(\bigoplus_{n=1}\Phi_{i}^{\left(n\right)}\left(\left\{ F_{j} \right\} \right)\Big) \cdot W
\end{equation}
where 
$\Phi_{i}^{\left(n\right)}\left(\left\{ F_{j}\right\} \right)$ 
is a general covariant $n$-atom interaction term for the activations $\left\{ F_{j}\right\}_{j \in \mathcal{S}_i}$.

The $n$-atom interactions for atom $i$ are constructed by summing over all possible paths $i \rightarrow j_{1}\rightarrow\ldots\rightarrow j_{n}$ of length $n$ which start atom $i$ and jump between atoms $j_k \in \mathcal{S}_i$. For each step in the path, we CG-multiply by $F_{j_k}$, along with a $\SO(3)$-vector transition ``amplitude'' $\Upsilon^{(n)}_{jj'}$:
\begin{equation}
\Phi_{i}^{\left(n\right)}\left(\left\{ F_{j} \right\} \right)=
\bigoplus_{m=0}^n
\bigoplus_{\{i_1, \ldots, i_m \} \subset \{1, \ldots, n\}}
\sum_{\substack{{j_1, \ldots, j_n} \in \mathcal{S}_i \\
j_{i_1} = \ldots = j_{i_m}=i}}
\bigotimes_{k=0}^{n-1}\left(\Upsilon_{j_{k}j_{k+1}}^{\left(n\right)}\otimes F_{j_{k+1}}\right)
\end{equation}
This form ensures that interactions are permutation invariant, translation invariant, 
and rotationally covariant. 
Here, we focus on a few key points:
%
\begin{enumerate}[~~1.]
\item We are building a representation of atom $i$ and atoms $j \in \mathcal{S}_{i}$ in its local environment. We therefore use the direct sum when $j_k = i$, and a normal sum otherwise.   
\item The form of  $\Upsilon^{(n)}_{jj^\prime}$ can be different for each $n$, and is constrained by symmetry, and unless otherwise noted we chose $\Upsilon_{jj} = 1$. 
\item If we require $\Upsilon^{(n)}_{jj^\prime}$ depend only on the relative position $\mathbf{r}_{jj'} = \mathbf{r}_j - \mathbf{r}_{j'}$ for $j \neq j^\prime$, then $\Upsilon^{(n)}_{jj^\prime}$ must be constructed only from linear combinations of tensor powers of $\mathbf{r}_{jj'}$. A decomposition of $\bigoplus_k\mathbf{r}_{jj'}^{\otimes k}$ into irreducibles generates terms proportional to $r_{jj'}^m Y^\ell(\hat{\mathbf{r}}_{jj^\prime})$. 
We therefore chose the generic form $\Upsilon^{(n)}_{jj^\prime} = \Upsilon^{(n)}(\mathbf{r}_{jj^\prime}) = \bigoplus_{\ell=0}^{L} \mathcal{F}^\ell(r_{ij}) Y^\ell(\hat{\mathbf{r}}_{jj^\prime})$, where $\mathcal{F}_c^\ell(r)$ are a set of (possibly learnable) radial basis functions. 
\item The one-body interaction $\Phi_{i}^{\left(1\right)} = F_i \oplus \big(\sum_j \Upsilon^{(1)}_{ij} \otimescg F_j \big)$ contains a component analogous to the radial filters of 
\citep{Thomas2018}. 
\item Some of the information in higher order order interactions is induced from lower order terms. For example, $\Phi_{i}^{\left(2\right)} = \big( \Phi_{i}^{\left(1\right)} \big)^{\otimescg 2} \oplus \tilde{\Phi}_{i}^{\left(2\right)}$, where $\tilde{\Phi}_{i}^{\left(2\right)}$ only contains terms of the form $F_{j_1} \otimescg F_{j_2}$, where $j_1 \neq j_2 \neq i$. 
\end{enumerate}
%
We use the $n$-atom interactions $\Phi_{i}^{\left(n\right)}$ to design the CG layers in our \emph{Cormorant} network. 
Note that while these can be structurally identified with $n$-atom interactions, we encourage the reader to not take the analogy too far. Each CG layer serves two purposes: (1) to build up a representation for an atom's local environment, and (2) to generate interactions between representations. Lower CG layers likely serve to build up a good representation. Only at higher layers, when the features $F_i$ are well constructed is it likely that a clear mapping to physical degrees of freedom be possible. We leave this connection to future work. 

\subsection{Implementation of $\textrm{SO(3)}$-vector layers}

The CG layers we chose for our implementation of Cormorant were based upon $n \leq 3$-atom interaction terms. Due to computational limitations, we considered only a subset of the general $n \leq 3$ form above. There is an analogy between these interactions and a generalization of Message Passing Neural Networks. In the supplement, we discuss the general form of the CG layer we implemented, along with the remaining details of our network architecture.
}

\subsection{Overall structure and comparison with other architectures}

In addition to the covariant neurons described above, our network also needs neurons to compute 
the input featurization and the the final output after the covariant layers. Thus, in total, 
a Cormorant networks consists of three distinct parts: 
\begin{compactenum}[~~1.]
\item An input featurization network $\{F^{s=0}_j\} \leftarrow \mathrm{INPUT}(\{Z_i, r_{i,j}\})$ that 
operates only on atomic charges/identities and (optionally) a scalar function of relative positions $r_{i,j}$. 
\item An $S$-layer network  $\{ F^{s+1}_i \} \leftarrow \mathrm{CGNet}(\{ F^s_i \})$of covariant activations 
$F^{s}_i$, each of which is a $\SO(3)$-vector of type $\tau^s_i$. 
\item A rotation invariant network at the top $y \leftarrow \mathrm{OUTPUT}(\bigoplus_{s=0}^{S} \{F^s_i\})$ 
that constructs scalars from the activations $F^s_i$, and uses them to predict a regression target $y$. 
\end{compactenum}
We leave the details of the input and output featurization to the Supplement. 

A key difference between Cormorant and other recent covariant networks (Tensor Field Networks~\citep{Thomas2018} and \m{\textrm{SE}(3)}-equivariant networks~\citep{Weiler}) is the use of Clebsch-Gordan non-linearities. The Clebsch-Gordan non-linearity results in a complete interaction of every degree of freedom in an activation. This comes at the cost of increased difficulty in training, as discussed in the Supplement.  We further note that \m{\textrm{SE}(3)}-equivariant networks use a three-dimensional grid of points to represent data, and ensure both translational and rotational covariance (equivariance) of each layer. Cormorant on the other hand uses activations that are covariant to rotations, and strictly invariant to translations.


\section{Experiments}

We present experimental results on two
datasets of interest to the computational chemistry community: MD-17
for learning molecular force fields and potential energy surfaces,
and QM-9 for learning the ground state properties of a set of molecules. 
The supplement provides a detailed summary of all hyperparameters, our training algorithm, and the details of the input/output levels used 
in both cases. Our code is available at \href{https://github.com/risilab/cormorant}{https://github.com/risilab/cormorant}.

\textbf{QM9}~\citep{ramakrishnan2014quantum} is a dataset of approximately \m{134}k small organic molecules 
containing the atoms H, C, N, O, F.~ 
For each molecule, the ground state configuration is calculated using DFT, along with a variety of molecular properties. 
We use the ground state configuration as the input to our Cormorant, 
and use a common subset of properties in the literature as regression targets. 
Table~\ref{tab:results}(a) presents our results averaged over three training runs compared with 
SchNet~\citep{SchNet}, MPNNs~\citep{Riley2017}, and wavelet scattering networks~\citep{Hirn2017}. 
Of the twelve regression targets considered, we achieve leading or competitive results on six 
($\alpha$, $\Delta\epsilon$, $\epsilon_{\mathrm{HOMO}}$, $\epsilon_{\mathrm{LUMO}}$, $\mu$, $C_v$).  
The remaining four targets are within $40\%$ of the best result, with the exception of $R^2$.

\textbf{MD-17}~\citep{Chmiela2016a} is a dataset of eight small organic molecules
(see Table~\ref{tab:results}(b)) containing up to 17 total atoms composed of the atoms H, C, N, O, F.~  
For each molecule, an \emph{ab initio} molecular dynamics simulation was run using DFT
to calculate the ground state energy and forces. At intermittent timesteps,
the energy, forces, and configuration (positions of each atom) were
recorded. For each molecule we use a train/validation/test split of
50k/10k/10k atoms respectively. 
The results of these experiments are presented in Table~\ref{tab:results}(b), where the
mean-average error (MAE) is plotted on the test set for each of molecules. (All units are in kcal/mol, as consistent with the dataset and the literature.)
To the best of our knowledge, the current state-of-the art algorithms on this dataset are 
DeepMD~\citep{Zhang2017}, DTNN~\citep{Schutt2017}, SchNet~\citep{SchNet}, GDML~\citep{Chmiela2016a}, and sGDML~\citep{Chmiela2018}.
Since training and testing set sizes were not consistent, we
used a training set of 50k molecules to compare with all neural network
based approaches. As can be seen from the table, our \emph{Cormorant} network
outperforms all competitors. 

\begin{table}[t]
\centering
\caption{\label{tab:results} Mean absolute error of various prediction targets on QM-9 (left) 
and conformational energies (in units of kcal/mol) on MD-17 (right). The best results within a standard deviation of three Cormorant training runs (in parenthesis) are indicated in bold.}
\begin{minipage}{0.49\textwidth}
\tiny
\setlength\tabcolsep{3pt}
\begin{tabular}{lrrrrr}
\toprule
{} &  Cormorant & &  SchNet &     NMP &  WaveScatt \\
\midrule
$\alpha$ ($\mathrm{bohr}^3$) &\textbf{      0.085}&(0.001)&   0.235 &   0.092 &      0.160 \\
$\Delta \epsilon$ (eV)       &\textbf{      0.061}&(0.005)&\textbf{   0.063}&   0.069 &      0.118 \\
$\epsilon_{\rm HOMO}$ (eV)   &\textbf{      0.034}&(0.002)&   0.041 &   0.043 &      0.085 \\
$\epsilon_{\rm LUMO}$ (eV)   &\textbf{      0.038}&(0.008)&\textbf{   0.034}&\textbf{   0.038}&      0.076 \\
$\mu$ (D)                    &\textbf{      0.038}&(0.009)&\textbf{   0.033}&\textbf{   0.030}&      0.340 \\
$C_v$ (cal/mol K)            &\textbf{      0.026}&(0.000)&   0.033 &   0.040 &      0.049 \\
$G$ (eV)                     &      0.020 &(0.000)&\textbf{   0.014}&   0.019 &      0.022 \\
$H$ (eV)                     &      0.021 &(0.001)&\textbf{   0.014}&   0.017 &      0.022 \\
$R^2$ ($\mathrm{bohr}^2$)    &      0.961 &(0.019)&\textbf{   0.073}&   0.180 &      0.410 \\
$U$ (eV)                     &      0.021 &(0.000)&\textbf{   0.019}&   0.020 &      0.022 \\
$U_0$ (eV)                   &      0.022 &(0.003)&\textbf{   0.014}&   0.020 &      0.022 \\
ZPVE (meV)                   &      2.027 &(0.042)&   1.700 &\textbf{   1.500}&      2.000 \\
\bottomrule
\end{tabular}

\end{minipage}
\begin{minipage}{0.49\textwidth}
\tiny
\setlength\tabcolsep{3pt}
\begin{tabular}{lrrrrrr}
\toprule
{} &  Cormorant &  DeepMD &  DTNN &  SchNet &  GDML &  sGDML \\
\midrule
Aspirin        &\textbf{      0.098 }&   0.201 &    -- &   0.120 & 0.270 &  0.190 \\
Benzene        &\textbf{      0.023 }&   0.065 & 0.040 &   0.070 & 0.070 &  0.100 \\
Ethanol        &\textbf{      0.027 }&   0.055 &    -- &   0.050 & 0.150 &  0.070 \\
Malonaldehyde  &\textbf{      0.041 }&   0.092 & 0.190 &   0.080 & 0.160 &  0.100 \\
Naphthalene    &\textbf{      0.029 }&   0.095 &    -- &   0.110 & 0.120 &  0.120 \\
Salicylic Acid &\textbf{      0.066 }&   0.106 & 0.410 &   0.100 & 0.120 &  0.120 \\
Toluene        &\textbf{      0.034 }&   0.085 & 0.180 &   0.090 & 0.120 &  0.100 \\
Uracil         &\textbf{      0.023 }&   0.085 &    -- &   0.100 & 0.110 &  0.110 \\
\bottomrule
\end{tabular}

\end{minipage}

\end{table}

\section{Conclusions}

To the best of our knowledge, \emph{Cormorant} is the first neural network architecture in which 
the operations implemented by the neurons is directly motivated by the form of known physical interactions. 
Rotation and translation invariance are explicitly ``baked into'' the network by the fact all activations 
are represented in spherical tensor form (\m{\SO(3)}--vectors), and the neurons combine Clebsch--Gordan 
products, concatenation of parts and mixing with learnable weights, all of which are  covariant operations. 
In future work we envisage the potentials learned by \emph{Cormorant} to be directly integrated in 
MD simulation frameworks.
In this regard, it is very encouraging that on MD-17, which is the standard benchmark for force field 
learning, \emph{Cormorant} outperforms all other competing methods. 
Learning from derivatives (forces) and generalizing to other compact symmetry groups are 
natural extensions of the persent work. 

\subsection*{Acknowledgements}

This project was supported by DARPA ``Physics of AI'' grant number HR0011837139, and used computational 
resources acquired through NSF MRI 1828629. 

We thank E. Thiede for helpful discussion and comments on the manuscript.

\ignore{
We propose a new architecture, we call Cormorant, for learning on molecular data from DFT calculations. 
These networks provide a physics inspired platform for learning functions on molecular data. 
At the heart of the networks are $n$-atom interactions, that combine $n$ atomic representations using 
Clebsch-Gordan operations to covariantly construct a new atom representation. 
We consider a specific choice of interactions in this framework, and relate the corresponding architecture 
to a generalization of message passing neural networks. 
We train our Cormorant on two standard datasets in the molecular chemistry community. 
We find that for the problem of learning potential energy surfaces, we significantly outperform competing architecutres. For the task of learning ground state molecular properties we are competitive with the state of the art on many learning targets.
}

\ignore{
\clearpage 

Our network is constructed in three components: (1) An input featurization network $\{F^{s=0}_j\} \leftarrow \mathrm{INPUT}(\{Z_j, r_{jj'}\})$ that operates only on atomic charges/identities and (optionally) a scalar function of relative positions $r_{jj'}$. (2) An $S$-layer network  $\{ F^{s+1}_j \} \leftarrow \mathrm{CGNet}(\{ F^s_j \})$of covariant activations $F^{s}_i$, each of which is a $\SO(3)$-vector of type $\tau_i$. (3) A rotation invariant network at the top $y \leftarrow \mathrm{OUTPUT}(\bigoplus_{s=0}^{S} \{F^s_i\})$ that constructs scalars from the activations $F^s_i$, and uses them to predict a regression target $y$. In the following section we focus on the network that constructs the covariant activation functions, and leave the details of the input and output featurization to the Supplement. 

\subsection{Clebsch-Gordan non-linearity and \m{\SO(3)}-vector operations}
\label{sec:cg_nonlinearity}

The central operation in our Cormorant is the Clebsch-Gordan transformation applied to two $\SO(3)$ vectors $F_1$ and $F_2$, with types  $\tau_1 = \big(\tau_1^0, \ldots, \tau_1^{L} \big)$ and $\tau_2 = \big(\tau_2^0, \ldots, \tau_2^{L} \big)$. This requires a generalization of the transformation in Sec.~\ref{sec: physical}, defined for single component irreducible $\SO(3)$-vectors $Q^{A}_{\ell}$ and $Q^{B}_{\ell^\prime}$. 
The general form of the CG decomposition 
results in a quadratic increase in the number of parts. 
Here, we specialize to the case where $\tau_1^\ell = \tau_2^\ell = N_c$, where $N_c$ is the number of channels. Given this restriction, we henceforth define the CG decomposition between two $\SO(3)$-vectors as:
\begin{equation}
F_{1} \otimescg F_{2} = 
\bigoplus_{c}\bigoplus_{\ell = |\ell_{1} - \ell_{2}|}^{\ell_{1}+\ell_{2}}
C_{\ell_{1},\ell_{2},\ell} \cdot\left(F_{1,c}^{\ell}\otimes F_{2,c}^{\ell}\right).
\end{equation}
This structure is strictly less general than the form used in \citep{KLT2018}, as it takes the elements $c = c^\prime$ of the ``part'' indices. However, it is more computationally tractable, and no less expressive when combined with linear mixing matrices.

Throughout this text, $\oplus$ denotes the sum of $\SO(3)$-vectors, which concatenates the irreps of each isotypic part of both $\SO(3)$-vectors type $\tau_1$ and $\tau_2$ into a new $\SO(3)$-vector of type $\tau^\ell_3 = \tau^\ell_1 + \tau^\ell_2$. We also can mix $\SO(3)$-vectors component wise with a list of weight mixing matrices $W = (W^0, \ldots, W^\ell)$, which we denote through $\tilde{F} =  F \cdot W = \oplus_{\ell=0}^{L} F^\ell W^\ell$. It is useful to note that the CG product is only associative or commutative up to a unitary transformation, i.e., $\left( F_1 \otimescg F_2 \right) = \left( F_2 \otimescg F_1 \right) \cdot U$, and $\left( F_1 \otimescg F_2 \right) \otimescg F_3 = F_1 \otimescg \left( F_2 \otimescg F_3 \right) \cdot V$, for some set of unitary matrices $U_\ell U_\ell^\dagger = V_\ell V_\ell^\dagger = \hat{1}$. At times we will reabsorb these unitary matrices into a redefinition of the learnable weights $W$.
}

\clearpage 
\bibliographystyle{plainnat}
{\small
\setlength{\bibsep}{2pt}
\bibliography{Cormorant}
}



\clearpage

\setcounter{equation}{0}
\setcounter{figure}{0}
\setcounter{table}{0}
\setcounter{page}{1}
\makeatletter
\renewcommand{\theequation}{S\arabic{equation}}
\renewcommand{\thefigure}{S\arabic{figure}}
\renewcommand{\bibnumfmt}[1]{[S#1]}
\renewcommand{\citenumfont}[1]{S#1}



\section{Architecture}

As discussed in the main text, our Cormorant architecture is constructed from three basic building blocks: (1) an input featurization that takes $(Z_i, \mathbf{r}_i)$ and outputs a scalar, (2) a set of covariant CG layers that update $F_i^{s}$ to $F_i^{s+1}$, (3) a layer that takes the set of covariant activations $F_i^{s}$, and construct a permutation and rotation invariant regression target.

\subsection{Notation}

Throughout this section, we will follow the use the main text, and denote a $\textrm{SO(3)}$-vector at layer $s$ by $F^{s} = (F^{s}_{0}, \ldots, F^{s}_{L})$ with maximum weight $L$. Each $\textrm{SO(3)}$-vector has corresponding type $\tau^{s}$, and lives in a representation space $F^{s} \in V^{s} =\bigoplus_{\ell=0}^{L^{s}} \bar{V}_{\ell}^{\tau_{\ell}^{s}}$, where $\bar{V}_{\ell} = \mathbb{C}^{(2\ell+1) \times 1}$ is the representation space for irreducible representation of $\textrm{SO(3)}$ with multiplicity 1. We will also introduce the vector space for the edge network $V_{\rm edge} ^s= \bigoplus_{\ell=0}^{L^{s}} \mathbb{C}^{\tau_{\ell}^{s}}$. 

See Table~\ref{tab:symbols} for a more complete table of symbols used in the supplement and main text.

\subsection{Overall structure}

The Cormorant network is a function ${\rm CORMORANT}\left(\left\{ Z_{i},\mathbf{r}_{i}\right\} \right):\mathbb{Z}^{N}\times\mathbb{R}^{N\times3}\rightarrow\mathbb{R}$ that takes a set of $N$ charge-positions $\left\{ Z_{i},\mathbf{r}_{i}\right\} $ and outputs a single regression target. The 
\begin{equation}
{\rm CORMORANT}\left(\left\{ Z_{i},\mathbf{r}_{i}\right\} \right)={\rm OUTPUT}\left({\rm CGNet}\left({\rm INPUT}\left(\left\{ Z_{i},\mathbf{r}_{i}\right\} \right)\right)\right)
\end{equation}
 networks are constructed from three basic units:
\begin{enumerate}

\item ${\rm INPUT}\left(\left\{ Z_{i},\mathbf{r}_{i}\right\} \right):\mathbb{Z}^{N}\times\mathbb{R}^{N\times3}\rightarrow (\bar{V}_0)^N$ which takes the $N$ charge-position  pairs and outputs $N$ sets of scalar feature vectors $c_{{\rm in}}$. (See section \ref{sec:input_featurization}.)

\item ${\rm CGNet}\left(\left\{ F_i,\mathbf{r}_{i}\right\} \right):(\bar{V}_0)^N \times\mathbb{R}^{N\times3}\rightarrow\bigoplus_{s=0}^{S} \left( V^{s} \right)^{N}$ takes the set of scalar features from ${\rm INPUT}\left(\left\{ Z_{i},\mathbf{r}_{i}\right\} \right)$, along with the set of positions for each atom, and outputs a $\textrm{SO(3)}$-vector for each level $s = 0, \ldots, S$ using Clebsch-Gordan operations. (See section \ref{sec:covariant_layers}.)

\item ${\rm OUTPUT}\left(\bigoplus_{s=0}^{S} ( V^{s} )^{N} \right)\rightarrow\mathbb{R}$ takes the output of ${\rm CGNet}$ above, constructs a set of scalars, and then constructs a permutation-invariant prediction that can be exploited at the top of the network. (See section \ref{sec:output_featurization}.)

\end{enumerate}
This design is organized in a modular way to separate the input featurization, the covariant $\textrm{SO(3)}$-vector layers, and the output regression tasks. Importantly, the ${\rm INPUT}$ and ${\rm OUTPUT}$ networks are different for GDB9 and MD17. However, the covariant $\textrm{SO(3)}$-vector layers ${\rm CGNet}$ were identical in design and hyperparameter choice. We include these designs and choices below.


\subsection{Input featurization}
\label{sec:input_featurization}

\subsubsection{MD-17}

For MD-17, the input featurization was determined by taking the tensor product $\tilde{F}_i = \mathrm{onehot}_i \otimes \vec{Z}_i$, where $\mathrm{onehot}_i$ is a one-hot vector determining which of $N_{\rm species}$ atomic species an atom is, and $\vec{Z}_i = (1, \tilde{Z}_i, \tilde{Z}_i^2)$, where $\tilde{Z}_i = Z_i / Z_{\rm max}$, and $Z_{\rm max}$ is the largest charge in the dataset. We then use a single learnable mixing matrix to convert this real vector with $3\times N_{\rm species}$ elements to a complex representation $\ell = 0$ and $N_c$ channels (or $\tau_i = (n_c)$.) 

We found for MD-17, a complex input featurization network was not significantly beneficial, and that this input parametrization was sufficiently expressive. 

\subsubsection{QM-9}

For the dataset QM-9, we used an input featurization based upon message passing neural networks. We start by creating the vector $\tilde{F}_i = \mathrm{onehot}_i \otimes \vec{Z}_i$ as defined in the previous section. Using this, a weighted adjacency matrix is constructed using a mask in the same manner as in the main text: $\mu_{ij} = \sigma((r_{\rm cut} - r_{ij}) / w)$, with learnable cutoffs/width $r_{\rm cut}$/$w$ and $\sigma(x) = 1 / (1 + \exp(-x))$ . This mask is used to aggregate neighbors $\tilde{F}^{\rm agg}_i = \sum_j \mu_{ij} \tilde{F}_j$. The result is concatenated with $\tilde{F}$, and passed through a MLP with a single hidden layer with 256 neurons and $\mathrm{ReLU}$ activations with an output real vector of length $2\times n_c$. This is then resized to form a complex $\SO(3)$-vector composed of a single irrep of type $\tau_i = (n_c)$.

\subsection{Covariant $SO(3)$-vector layers}
\label{sec:covariant_layers}

For both datasets, the central covariant $SO(3)$-vector layers of our Cormorant are identical. In both cases, we used $S = 4$ layers with $L = 3$, followed by a single $SO(3)$-vector layer with  $L = 0$. The number of channels of the input tensors at each level is fixed to $n_c = 16$, and similarly the set of weights $W$ reduce the number of channels of each irreducible representation back to $n_c = 16$.

\subsubsection{Overview}

The algorithm can be implemented as iterating over the function 
$${\rm CGLayer}\left(g_{ij}^{s},F_{i}^{s},\mathbf{r}_{i,}\right): (V^{s}_{\rm edge})^{N\times N} \times\mathbb{R}^{N\times N\times 3} \times (V^s)^{N}\rightarrow(V^{s+1}_{\rm edge})^{N\times N} \times(V^{s+1})^{N}$$
where $g_{ij}^{s}\in (V_{\rm edge}^s)^{N \times N}$ and is an edge network at level $s$ with $c_{s}$ channels for each $\ell\in\left[0,L\right]$, and $F_{i}^{s}\in (V^s)^{N}$ is an atom-state vector that lives in the representation space at level $s$. 

The function $\left(g_{ij}^{s+1}, F_{i}^{s+1}\right)\leftarrow{\rm CGLayer}\left(g_{ij}^{s},F_{i}^{s},\mathbf{r}_{i}\right)$ is itself constructed in the following way:
\begin{itemize}
\item $g_{ij}^{s+1} \leftarrow{\rm EdgeNetwork}\left(g_{ij}^{s},\mathbf{r}_{ij},F_{i}^{s}\right)$
\item $G^{s+1}_{ij}\leftarrow{\rm Edge2Vertex}\left(g_{ij}^{s+1},Y^{\ell}\left(\hat{\mathbf{r}}_{ij}\right)\right)$
\item $F_{i}^{s+1}\leftarrow{\rm VertexNetwork}\left(F^{s+1}_{ij},F_{i}^{s}\right)$
\end{itemize}
where:
\begin{enumerate}
\item ${\rm EdgeNetwork}\left(g_{ij}^{s},\mathbf{r}_{ij},F_{i}^{s}\right):(V^{s}_{\rm edge})^{N\times N}\times\mathbb{R}^{N\times N\times3}\times(V^{s})^{N}\rightarrow(V^{s+1}_{\rm edge})^{N\times N}$ is a pair/edge network that combined the input pair matrix $g_{ij}^{s}$ at level $s$, with a position network $F_{ij,c}=F_{c}\left(\left|\mathbf{r}_{ij}\right|\right)$, and $d_{ij} \sim F_{i} \cdot F_{j}$ is a matrix of dot products, all of which will be defined below. This output is then used to construct a set of representations that will be used as the input to the ${\rm VertexNetwork}$ function below.

\item ${\rm Edge2Vertex}\left(g_{ij}^{s+1},Y_{ij}\right):(V^{s}_{\rm edge})^{N\times N}\times (V)^{N \times N}\rightarrow (V^s)^{N\times N}$ takes the product of the scalar pair network $g_{ij}^{s+1}$, with the $SO(3)$-vector of spherical harmonics $Y_{ij} = \bigoplus_{\ell=0}^{L}Y^{\ell}\left(\hat{\mathbf{r}}_{ij}\right)$, to produce a $SO(3)$-vector of edge scalar representations that will be considered in the aggregation step in ${\rm VertexNetwork}$.

\item ${\rm VertexNetwork}\left(G^{s+1}_{ij},F_{i}^{s}\right):(V^s)^{N \times N} \times(V^s)^{N}\rightarrow (V^{s+1})^{N}$ updates the vertex $\textrm{SO(3)}$-vector activations by combining a ``Clebsch-Gordan aggregation'', a CG non-linearity, a skip connection, and a linear mixing layer.
\end{enumerate}

\subsubsection{Edge networks}

Our edge network is an extension of the ``edge networks'' in Message Passing Neural Networks ~\cite{Gilmer2017}. The $\rm EdgeNetwork$ function takes three different types of pair features, concatenates them, and then mixes them. We express write the edge network (Eq.~(9)) in the main text) with all indices explicitly included:
\begin{equation}
g_{\ell c,ij}^{s+1}=m_{c,ij}^{s}\odot\sum_{c^{\prime}} \left(\bigoplus_{c_1} g_{\ell c_{1},ij}^{s}\oplus \bigoplus_{c_2} d_{c_{2},ij}^{s}\oplus \bigoplus_{c_3}  \eta_{\ell c_{3},ij}\right)_{c^{\prime}} \left(W_{s,\ell}^{\rm edge} \right)_{c'c}
\end{equation}
where:
\begin{itemize}
\item $W_{s,\ell}^{\rm edge}$ is a weight matrix at layer $s$ for each $\ell$ of the edge network.

\item $g_{\ell c_{1},ij}^{s}$ is a set of edge activations from the previous layer.

\item $d_{c_{2},ij}^{s}=\bigoplus_{\ell=0}^{L}F^s_{\ell c_2 i}\cdot F^s_{\ell c_2 j}$, is a matrix of dot products, where $F_{\ell ci}\cdot F_{\ell cj}=\sum_{m}\left(-1\right)^{m}\left(F_{\ell ci,m}F_{\ell cj,-m}\right)$.\footnote{Note that $F_{\ell ci}\cdot F_{\ell cj}=\sum_{m}\left(-1\right)^{m}\left(F_{\ell ci,m}F_{\ell cj,-m}\right)$ is (up to a constant) just the CG decomposition $C_{\ell\ell0}\left(F_{\ell ci}\otimes F_{\ell cj}\right)$. The specific matrix elements of the CG coefficients $C_{\ell\ell0}$ are $\left\langle \ell m_{1}\ell m_{2}|00\right\rangle \propto\left(-1\right)^{m_{1}}\delta_{m_{1},-m_{2}}$.}

\item $\eta_{\ell c_{3},ij}^{s}=\eta_{\ell c_3}^{s}\left(\left|\mathbf{r}_{ij}\right|\right)$ is a set of learnable basis functions. These functions are of the form $\eta_{\ell c_{k,n}}^{s}\left(r\right)=r^{-k}\left(\sin\left(2\pi\kappa_{\ell n}^{s}r+\phi_{\ell n}^{s}\right) + \mathrm{i} \sin\left(2\pi\bar\kappa_{\ell n}^{s} r+\bar\phi_{\ell n}^{s}\right)\right)$, where $\kappa_{\ell n}^{s}$,  $\bar\kappa_{\ell n}^{s}$, $\phi_{\ell n}^{s}$, and $\bar\phi_{\ell n}^{s}$ are learnable parameters, the list of channels $c$ is found by flattening the matrix indexed by $c_{3}=\left(k,n\right)$, and $\mathrm{i}^2 = -1$. 

\item $\mu_{\ell c,ij}^{s}$ is a mask that is used drop the radial functions smoothly to zero. This mask is constructed through $$\mu_{c,ij}=\sigma\left(-\left(r_{ij}-r_{c,{\rm soft}}^{s}\right)/w_{c}^{s}\right),$$ where $\sigma\left(x\right)$ is the sigmoid activation, $r_{c,{\rm soft}}^{s}$ is a soft cutoff that drops off with width $w_{c}^{s}$. 
\end{itemize}

\subsubsection{From edge scalar representations to $SO(3)$-vector}

The function $G^{s+1}_{ij}\leftarrow{\rm Edge2Vertex}\left(g_{ij}^{s+1},Y^{\ell}\left(\hat{\mathbf{r}}_{ij}\right)\right)$ will take the scalar output of the edge network $g_{\ell c,ij}^{s+1}$, and construct a set of $SO(3)$-vector representations using spherical harmonics through:
\begin{equation}
G^{s+1}_{\ell c,ij}=g_{\ell c,ij}^{s}Y^{\ell}\left(\hat{\mathbf{r}}_{ij}\right)
\end{equation}
We note the normalization of the spherical harmonics here is not using the ``quantum mechanical'' convention, but rather are normalized such that $\sum_{m}\left|Y_{m}^{\ell}\left(\hat{\mathbf{r}}\right)\right|^{2}=1$. This is equivalent to scaling the QM version by $Y_{m}^{\ell}\left(\hat{\mathbf{r}}\right)\rightarrow\sqrt{\frac{2\ell+1}{4\pi}}\times Y_{m}^{\ell}\left(\hat{\mathbf{r}}\right)$.

\subsubsection{Vertex networks}

The function ${\rm VertexNetwork}$ is found by concatenating three operations:
\begin{align}
F_{\ell,i}^{s+1} & = \left( {\rm VertexNetwork}\left(G^{s+1}_{ij},F_{i}^{s}\right) \right)_\ell\\
 & = \sum_{c^{\prime}} \left(\bigoplus_{c_1} F_{c_1,i}^{s+1,{\rm ag}}\oplus \bigoplus_{c_2} F_{c_2,i}^{s+1,{\rm nl}}\oplus \bigoplus_{c_3} F_{c_3,i}^{s+1,{\rm id}}\right)_{\ell,c^{\prime}} \left(W^{{\rm vertex}}_{s,\ell}\right)_{c^{\prime}c}
\end{align}
 where
\begin{enumerate}
\item $F_{i}^{s+1,{\rm ag}}=\sum_{j\in N\left(i\right)} G^{s+1}_{ij} \otimescg F_{j}^{s} $ is a CG-aggregation step and $G^{s}_{ij}$ is the set of edge representations calculated by ${\rm Edge2Vertex}$.

\item $F_{i}^{s+1,{\rm nl}}=F_{i}^{s} \otimescg F_{i}^{s}$
is a CG non-linearity.

\item $F_{i}^{s+1,{\rm id}}=F_{i}^{s}$ is just the identity function, or equivalently a skip connection.

\item $W^{{\rm vertex}}_{s,\ell}$ is a atom feature mixing matrix. 
\end{enumerate}


\subsection{Output featurization}
\label{sec:output_featurization}

The output featurization of the network starts with the construction of a set of scalar invariants from the set of activations $F_i^{s}$ for all atoms $i$ and all levels $s=0\ldots S$. We extract three scalar invariants from each activation $F$ (dropping the $i$ and $s$ indices):
\begin{compactenum}
\item Take the $\ell=0$ component: $\xi_0(F) = F^{s}_{\ell=0}$.
\item Take the scalar product with itself: $\xi_1(F) = \mathrm{Re}[\tilde{\xi}_1(F)] + \mathrm{Im}[\tilde{\xi}_1(F)]$ where $\tilde{\xi}_1(F) = \sum_{m=-\ell}^{\ell} (-1)^{m} F^s_{\ell,m} F^s_{\ell,-m}$.
\item Calculate the $\SO(3)$-invariant norm: $\xi_2(F^s) = \sum_{m=-\ell}^{\ell} F^s_{\ell,m} \left(F^s_{\ell,m}\right)^*$.
\end{compactenum}
These are then concatenated together to get a final set of scalars:
$x_i = \bigoplus_{s=0}^{S} \xi_0(F_i^s)  \oplus \bigoplus_{\ell=0}^{L} (\xi_1(F_i^s)  \oplus \xi_2(F_i^s))$
and fed into the output network network.

\subsubsection{MD-17}

The output for the MD-17 network is straightforward. The scalars $x_i$ are summed over, and then a single linear layer is applied: $y = A \left(\sum_i x_i \right) + b$. 

\subsubsection{QM-9}

The output for the QM-9 is constructed using two multi-layer perceptrons (MLPs). First, a MLP is applied to the scalar representation $x_i$ at each site. The result is summed over all sites, forming a single permutation invariant representation of the molecule. This representation is then used to predict a single number used as the regression target:
$y = \mathrm{MLP}_2 \left(\sum_i \mathrm{MLP}_1(x_i)\right)$. Here, both $\mathrm{MLP}_1$ and $\mathrm{MLP}_2$ have a single hidden layer of size 256, and the intermediate representation has 96 neurons.


\subsection{Weight initialization}
\label{sec:weight-initialization}

All CG weights $W^{\ell}$ were initialized uniformly in the interval $[-1, 1]$, and then scaled by a factor of $W^{\ell}_{\tau_\ell^{\rm in}, \tau_\ell^{\rm out}} \sim \mathrm{Unif}(-1, 1) * g / (\tau^{\rm in}_\ell + \tau^{\rm out}_\ell)$, where $\tau^{\rm in}_\ell$, $\tau^{\rm out}_\ell$  and $g$ is the weight gain. 

We chose the gain to ensure that the activations at each level were order unity when the network is initialized. We found that if the gain was too low, the CG products in higher levels would not significantly contribute to training, and information would only flow through linear (one-body) operations. This would result in convergence to poor training error. On the other hand, if the gain is set too high, the CG non-linearities dominate at initialization and would increase the change of the instabilities discussed above. 

In practice, the gain was hand-tuned by such that the mean of the absolute value of the CG activations $1/{(N_{\rm atom} N_{\rm c} (2\ell+1))}\sum_{\ell, i,c,m} |F^s_{\ell,i,c,m}|$ at each level was approximately unity for a random mini-batch. For experimental results presented here, we used a gain of $g=5$. 

\section{Experimental details}\label{sec:training-algorithm}

We trained our network using the AMSGrad~\citep{j.2018on} optimizer with a constant learning rate of $5\times 10^{-4}$ and a mini-batch size of $25$. We trained for 512 and 256 epoch respectively for MD-17 and QM-9. For each molecule in MD-17, we uniformly sampled 50k/10k/10k data points in the training/validation/test splits respectively. In QM-9 the dataset was randomly split to 100k molecules in the train set, with $10\%$ in the test set, and the remaining in the validation set. We removed the 3054 molecules that failed consistency requirements~\citep{ramakrishnan2014quantum}, and also subtracted the thermochemical energy~\citep{Gilmer2017} for the targets $C_v$, $U_0$, $U$, $G$, $H$, ZPVE.

For both datasets, we used, $S=4$ CGLayers with $L=3$ and we used $N_c = 16$ channels at the output of each CGLayer. This gave networks with 299808 and 154241 parameter respectively for QM-9 and MD-17. Training time for QM-9 is takes roughly 48 hours on a NVidia 2080 Ti GPU. Training time for MD-17 varies based upon the molecule being trained, but typically ranges between 26 and 30 hours.

\subsection{Training instabilities}

Training our Cormorant had several subtleties that we both believe are related to the nature of the CG non-linearity. 
We found a poor choice of weight initialization or optimization algorithm will frequently result in either: (1) an instability resulting in very large training loss ($> 10^6$), from which the network will never recover, or (2) convergence to weights where the activation of CG non-linearities in higher layers turn off, and the resulting training error is poor. 

We believe these difficulties are a result of the CG non-linearity, which is quadratic and unbounded. In fact, our network is just a high-order polynomial function of learnable parameters.\footnote{This is true for MD-17, although for QM-9, the presence of non-linearities in the fully-connected MLPs adds a more conventional non-linearity.}
For the hyperparameters used in our experiments, the prediction at the top is a sixteenth order polynomial of our network's parameters. As a result, in certain regions of parameter space small gradient updates can result in rapid growth of the output amplitude or a rapid drop in the importance of some channels.

These issues were more significant when we used Adam~\citep{Kingma2015AdamAM} then AMSGrad, and when the network's parameters were not initialized in a narrow range. Using the weight initialization scheme discussed in Sec.~\ref{sec:weight-initialization}, we were able to consistently converge to low training and validation error, provided we were limited to at most four CG layers.

\end{document}


\begin{table}[tb]
\begin{tabular}{|c|l|}
\hline 
Symbol & Meaning\tabularnewline
\hline 
\hline 
$Z_{i}$ & Atomic charge of atom $i$ (e.g., Hydrogen would be $Z_{i}=1$, and
Carbon would be $Z_{i}=6$)\tabularnewline
\hline 
$\mathbf{r}_{i}$ & Position of atom $i$\tabularnewline
\hline 
$\mathbf{r}_{ij}$ & Relative position of atoms $i$ and $j$: $\mathbf{r}_{ij}=\mathbf{r}_{i}-\mathbf{r}_{j}$\tabularnewline
\hline 
$\hat{\mathbf{r}}_{ij}$ & Unit vector in the direction of $\mathbf{r}_{ij}$\tabularnewline
\hline 
$s$ & Index of a ${\rm CGLayer}$ in Cormorant\tabularnewline
\hline 
$S$ & Top ${\rm CGLayer}$\tabularnewline
\hline 
$\ell$ & Index of an irreducible representation of ${\rm SO\left(3\right)}$\tabularnewline
\hline 
$L$ & Largest index in a representation\tabularnewline
\hline 
$m$ & Component $m\in[-\ell,\ell]$ of a ${\rm SO\left(3\right)}$-vector\tabularnewline
\hline 
$\tau^{s}$ & Vector of multiplicities of a ${\rm SO\left(3\right)}$-vector at
layer $s$\tabularnewline
\hline 
$n_{c}$ & Number of output channels of a ${\rm CGLayer}$\tabularnewline
\hline 
$\bar{V}_{\ell}$ & Representation space for irrep with index $\ell$ and multiplicity
$\tau_{\ell}=1$: $\bar{V}_{\ell}=\mathbb{C}^{\left(2\ell+1\right)\times1}$\tabularnewline
\hline 
$V^{s}$ & Representation space for ${\rm SO\left(3\right)}$-vector of type
$\tau^{s}$: $V^{s}=\bigoplus_{\ell=0}^{L}\bar{V}_{\ell}^{\tau_{\ell}^{s}}$\tabularnewline
\hline 
$V_{{\rm edge}}^{s}$ & Vector space for an scalar edge network at layer $s$: $V_{{\rm edge}}^{s}=\bigoplus_{\ell=0}^{L}\mathbb{C}^{\tau_{\ell}^{s}}$\tabularnewline
\hline 
$F_{i}^{s}$ & ${\rm SO\left(3\right)}$-vector atom-activation at layer $s$ for
atom $i$ \tabularnewline
\hline 
$Y^{\ell}\left(\hat{\mathbf{r}}_{ij}\right)$ & Spherical harmonics of weight $\ell$ in the direction of $\hat{\mathbf{r}}_{ij}$\tabularnewline
\hline 
$Y_{ij}$ & ${\rm SO\left(3\right)}$-vector of spherical harmonics connecting
atoms $i$ and $j$: $Y_{ij}=\bigoplus_{\ell=0}^{L}Y^{\ell}\left(\hat{\mathbf{r}}_{ij}\right)$\tabularnewline
\hline 
$g_{ij}^{s}$ & Scalar edge activation layer $s$ connecting atoms $i$ and $j$\tabularnewline
\hline 
$d_{ij}^{s}$ & Scalar matrix of dot products $d_{ij}^{s}=F_{i}^{s}\cdot F_{j}^{s}$\tabularnewline
\hline 
$\eta_{ij}^{s}$ & Radial basis functions between atoms $i$ and $j$\tabularnewline
\hline 
$G_{ij}^{s}$ & ${\rm SO\left(3\right)}$-vector edge activation layer $s$ connecting
atoms $i$ and $j$\tabularnewline
\hline 
$W_{s,\ell}^{{\rm vertex}}$ & Weight mixing matrix for component $\ell$ of atom ${\rm SO\left(3\right)}$-vector
activations at layer $s$\tabularnewline
\hline 
$W_{s,\ell}^{{\rm edge}}$ & Weight mixing matrix for component $\ell$ of scalar edge activations
at layer $s$\tabularnewline
\hline 
\end{tabular}

\caption{Table of symbols used in the main text and supplement.}
\end{table}